\documentstyle[aps,prd,preprint,floats,epsfig]{revtex}

\begin{document}
\draft

\tighten 
\preprint{\tighten\vbox{
                        \hbox{\hfil CLNS 99/1628}
                        \hbox{\hfil CLEO 99--8}
}}


\title{Structure Functions in the Decay $\tau^\mp \rightarrow \pi^\mp \pi^0\pi^0\nu_\tau$}  

\author{CLEO Collaboration}
\date{\today}

\maketitle
\tighten

\begin{abstract} 
Using the CLEO II detector operating at the CESR $e^+ e^-$ collider, 
we have measured the structure functions in the decay
$\tau^\mp \rightarrow \pi^\mp \pi^0\pi^0\nu_\tau$,
based on a sample corresponding to $4 \times 10^6$ produced $\tau$-pair events.
We determine the integrated structure functions, which depend only on the three
pion invariant mass, as well as the structure functions differential in the
Dalitz plot. 
We extract model independent limits on non-axial-vector contributions from the
measured structure functions as less than $16.6\%$ of the total branching fraction,
at the $95\%$ confidence level. 
Separating the non-axial-vector contributions into scalar and vector contributions,
we measure that scalars (vectors) contribute with less than $9.4\%$ ($7.3\%$)
to the total branching ratio, at the $95\%$ confidence level. 
\end{abstract}


\newpage

\begin{center}
T.~E.~Browder,$^{1}$ Y.~Li,$^{1}$ J.~L.~Rodriguez,$^{1}$
H.~Yamamoto,$^{1}$
T.~Bergfeld,$^{2}$ B.~I.~Eisenstein,$^{2}$ J.~Ernst,$^{2}$
G.~E.~Gladding,$^{2}$ G.~D.~Gollin,$^{2}$ R.~M.~Hans,$^{2}$
% ...
\end{center}
 
\small
\begin{center}
$^{1}${University of Hawaii at Manoa, Honolulu, Hawaii 96822}\\
$^{2}${University of Illinois, Urbana-Champaign, Illinois 61801}\\
% ...
\end{center}

\newpage


\section{INTRODUCTION}
\label{s-intro}

The hadronic structure in the decays 
$\tau^\mp\rightarrow\pi^\mp\pi^\mp\pi^\pm\nu_\tau$ and 
$\tau^\mp\rightarrow\pi^\mp\pi^0  \pi^0  \nu_\tau$ has been the
focus of several studies in recent 
years~\cite{argus1,argus2,delphi,cleo_3pi,opal}.  
In most of these studies, models are employed to characterize 
the hadronic structure.  As data samples have grown, revealing  
new and complex features, the models used have also become more 
complicated.  For example, simple models assume the $3\pi$ mass 
spectrum in these decays can be described by a single resonance, 
the $a_1(1260)$ meson.  
DELPHI \cite{delphi} found it necessary to include a radially 
excited $a_1$ meson, the $a^\prime_1$ meson.  Subsequently, 
CLEO \cite{cleo_3pi}, obtained a slightly improved description 
when including an $a_1^\prime$ meson, but found that the most 
significant improvement came about from accounting for 
a threshold effect in the mass-dependence of the $a_1$ width due to
the opening of the $a_1\to K^\star K$ decay channel.  
Similarly, models attributing structure in the $3\pi$ Dalitz 
plot distributions solely to decay amplitudes associated with 
$a_1\to \rho\pi$ were found to be insufficient by ARGUS \cite{argus2} 
and CLEO \cite{cleo_3pi}, who see indications of isospin zero 
contributions, from $f_2(1270)$, $f_0(1370)$ and $f_0(400-1200)$ 
({\sl e.g.}, $\sigma$) meson production.  The models used are not 
unique, as significant variations in their form and 
content can lead to similar features in the distributions of 
observable quantities.  In addition, no model has 
so far given a fully satisfactory description of the data.  

Thus, although the studies done so far have improved our 
knowledge, we still do not have a clear picture of the hadronic 
dynamics in $\tau$ decay to neutrino plus three pions.  
Furthermore, it is difficult to compare results from different 
experiments and draw additional conclusions since each analysis 
can only be interpreted in the context of the specific model(s) 
considered.  

J.~H.~K\"uhn and E.~Mirkes \cite{kuhn-struc} have constructed  
a convenient approach for model independent studies through the 
determination of structure functions.  In this approach, the 
kinematics of the three pions are studied in a particular 
reference frame.  In this frame, contributions from the production 
of $3\pi$ systems with different spin-parity quantum numbers 
are separated into different structure functions, 
expressed in a completely model independent fashion.  This 
approach also has the advantage that the features in the 
data are condensed into a compact form which can be used to test models,  
alone or in conjunction with data from other experiments.  
So far, only OPAL \cite{opal} has determined the structure 
functions, 
using the decay $\tau^\mp\rightarrow\pi^\mp \pi^\mp \pi^\pm \nu_\tau$. 
They obtained the first fully model independent measurement of the signed 
$\tau$ neutrino helicity as well as a model independent limit 
on non-axial-vector contributions.

Here, we present a measurement of the structure functions in the 
decay $\tau^\mp\rightarrow\pi^\mp\pi^0\pi^0\nu_\tau$, based on 
data collected with the CLEO~II detector corresponding to 
$4 \times 10^6$ produced $\tau$-pair events.  The sample of 
selected events is essentially the same as the one used in our 
model dependent analysis.  Consequently, we refer the reader to 
the article reporting on that work~\cite{cleo_3pi} for details 
common to both analyses.  

The outline for the 
rest of the article is as follows.  In the next section, we 
summarize the structure function formalism of K\"uhn and 
Mirkes~\cite{kuhn-struc} as applied to this decay mode.  
In Section~\ref{s-evtsel}, we describe the CLEO~II detector 
and the event selection criteria.  The method used to determine 
the structure functions is discussed in Section~\ref{s-method}, 
and the results are presented in Section~\ref{s-measure}.  The 
derived limits on non-axial-vector contributions are given in 
Section~\ref{s-nonaxial}.  In Section~\ref{s-sys} we describe
the sources of systematic error.  A summary and discussion of 
the results follows in Section~\ref{s-summ}.  

\section{THE STRUCTURE FUNCTION FORMALISM}
\label{s-strucintro}

Following the conventions of ref.~\cite{kuhn-struc}, we denote the
momenta of the three pions by $q_i^\mu$, with the momentum of the 
charged pion given by $q_3^\mu$ and the momenta of the two
neutral pions by $q^\mu_{1,2}$  
($\vert\vec{q}_2\vert > \vert\vec{q}_1\vert$). The momentum of the total
hadronic system is denoted by $Q^\mu = q_1^\mu + q_2^\mu + q_3^\mu$ and
its invariant mass by $\sqrt{Q^2}$. 

\begin{figure}[thb]
\begin{center}
\leavevmode
\unitlength1.0cm
\epsfxsize=8.cm
\epsfysize=8.cm
\epsffile{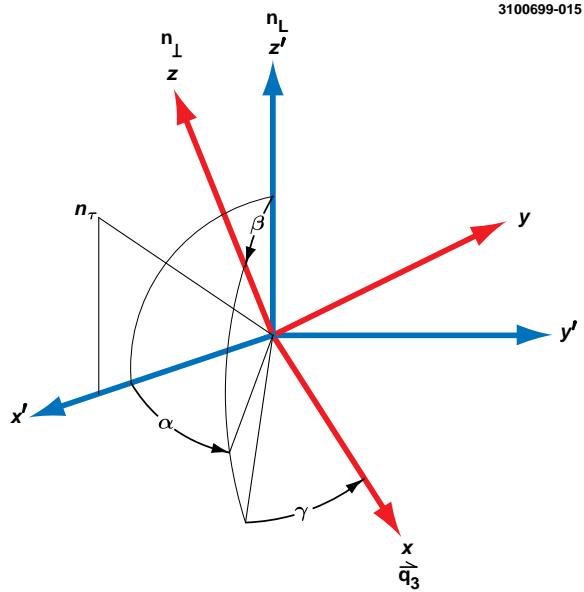}
\caption[]{\small Definition of the Euler angles $\alpha$, $\beta$, and $\gamma$
relating the two reference frames ${\cal S} (x,y,z)$ and 
${\cal S}^\prime (x^\prime , y^\prime , z^\prime )$, both defined in the 
three pion rest frame.}
\label{frame}
\end{center}
\end{figure}

For the description of the hadronic physics within the context of the
structure functions, it is convenient, as outlined in ref.~\cite{kuhn-struc},
to introduce two reference frames ${\cal S} (x,y,z)$ and 
${\cal S}^\prime (x^\prime, y^\prime , z^\prime )$, both defined in the three pion
rest frame. The two frames are transformed into each other via a rotation matrix
parametrized by the three Euler angles $\alpha$, $\beta$, and $\gamma$, as 
illustrated in Fig.~\ref{frame}. The orientation of the reference frame 
${\cal S} (x,y,z)$ is such that the $z$-axis is parallel to the normal 
$n_\perp = (\vec{q}_1 \times \vec{q}_2 ) / \vert \vec{q}_1 \times \vec{q}_2 \vert $
to the three pion decay plane and the $x$-axis is parallel to $\vec{q}_3$, the
flight direction of the third pion. The reference frame 
${\cal S}^\prime (x^\prime, y^\prime , z^\prime )$ has its $z^\prime$-axis aligned
with $n_L = - \vec{Q} / \vert \vec{Q} \vert $, {\it i.e.}~the flight 
flight direction of the laboratory frame as seen from the three pion
rest frame. Its azimuthal orientation is chosen such that $n_\tau$, the 
unobservable flight direction of the $\tau$ lepton in the three pion rest frame, 
lies in the $(x^\prime , z^\prime )$ plane.

In the ${\cal S} (x,y,z)$ reference frame, using the structure functions
$W_X$ and the corresponding lepton tensor combinations $\bar{L}_X$ of
K\"uhn and Mirkes \cite{kuhn-struc}, the differential decay rate for 
$\tau^\mp\rightarrow\pi^\mp\pi^0\pi^0\nu$ can 
be written as follows \cite{kuhn-struc}
\begin{eqnarray}
\label{eq:dgamma}
d\Gamma ( \tau\rightarrow 3\pi\nu_\tau ) &=& C(Q^2) 
\sum_X \bar{L}_X (\alpha ,\beta ,\gamma ,\theta_\tau ,Q^2 ) W_X (Q^2, s_1 , s_2 ) 
dQ^2 ds_1 ds_2 d\alpha d\gamma d\cos\beta d\cos\theta_\tau  
\nonumber \\  \rule[-2.0ex]{0pt}{6.0ex}
& & \mbox{ with } C(Q^2 ) =  \frac{G^2_F}{4 m_\tau (2\pi)^7  128  }  \cos^2\theta_c
\frac{(m^2_\tau - Q^2 )^2}{m^2_\tau Q^2}
\nonumber \\  \rule[-2.0ex]{0pt}{6.0ex} 
& & \mbox{ and } X \in \{A,B,\ldots, I,SA,SB,\ldots , SG \}   \, ,
\end{eqnarray}
where $G_F$ is the Fermi constant, $\theta_c$ the Cabibbo angle, and
$s_i$ the Dalitz plot variables with $s_i = (q_j + q_k )^2$ 
$(i,j,k=1,2,3; i\not= j\not= k )$. The Euler angles $\alpha$, $\beta$, 
and $\gamma$ as well as the invariant mass $\sqrt{Q^2}$ of the three pion 
system have been introduced above. The angle $\theta_\tau$ \cite{kuhn-struc}, which is 
the angle between the flight direction of the $\tau$ lepton in the 
laboratory system and the direction of the three pions in the 
$\tau$ lepton rest frame, is determined by the energy of the $\tau$ lepton 
in the laboratory frame, {\it i.e.}~approximately half the energy of the incoming
$e^+ e^-$ system, the energy of the three pion system in the laboratory frame,
and the invariant mass of the three pion system.
The sixteen structure functions $W_X$ and the corresponding
leptonic functions $\bar{L}_X$ are symmetric and antisymmetric combinations
of the hadron tensor $H^{\mu\nu}$ and, respectively, the lepton tensor 
$L^{\mu\nu}$. With the exception of the leptonic functions $\bar{L}_A$,
$\bar{L}_B$, and $\bar{L}_{SA}$ all leptonic functions $\bar{L}_X$ vanish
after integration over the angles $\alpha$, $\beta$, $\gamma$, 
and $\theta_\tau$. Accordingly, the decay rate in a given bin of 
$Q^2$ or $(Q^2 ,s_1 , s_2 )$ only depends on the structure functions 
$W_A$, $W_B$, and $W_{SA}$. For the details on the definitions of the 
structure functions $W_X$ as well as the leptonic functions $\bar{L}_X$  
the reader is referred to ref.~\cite{kuhn-struc}.

With the exception of the angle $\alpha$, which determines the azimuthal
orientation of the $\tau$ lepton, all variables in Eq.~\ref{eq:dgamma} are
measurable quantities in our experiment. Accordingly, in this analysis
we have to use the differential width $d\Gamma$ integrated over $\alpha$.
Performing this integral yields zero for the leptonic functions 
$\bar{L}_{SC}$, $\bar{L}_{SE}$, and $\bar{L}_{SG}$ (see ref.~\cite{kuhn-struc}). 
Thus, in our experiment the corresponding structure functions $W_{SC}$, $W_{SE}$, 
and $W_{SG}$ are not directly measurable.

Eq.~\ref{eq:dgamma} shows clearly the advantage of the chosen reference
frame ${\cal S} (x,y,z)$. The structure functions $W_X$ describing the 
hadronic physics depend on $Q^2$, $s_1$, and $s_2$, only. The remaining
dependence on the angular observables has been rotated into $\bar{L}_X$.
Accordingly, measurement of the set of structure functions 
$W_X(Q^2, s_1 , s_2 )$ in the reference frame ${\cal S} (x,y,z)$ 
yields a model idependent determination of the hadronic physics.

As mentioned above, the structure functions $W_X$ are symmetric and
antisymmetric combinations of the hadron tensor $H^{\mu\nu}$, which
on the other hand is derived from the hadronic current $h^\mu$ by
$H^{\mu\nu} = h^\mu h^{\star\nu}$. In general $h^\mu$ is given by
\cite{kuhn-struc}  
\begin{equation}
\label{eq:jmu}
h^\mu = V_1^\mu F_1 + V_2^\mu F_2 + i V_3^\mu F_3 + V_4^\mu F_4 \, ,
\end{equation}
where the form factors $F_i$ describe the unknown hadronic physics and the 
vectors $V_i^\mu$ the known spin structure. A possible representation is
\begin{eqnarray}
\label{eq:vmu}
V_1^\mu &=& q_1^\mu - q_3^\mu - Q^\mu \frac{ (Q(q_1 - q_3 ))}{Q^2} \nonumber\\
V_2^\mu &=& q_2^\mu - q_3^\mu - Q^\mu \frac{ (Q(q_2 - q_3 ))}{Q^2} \nonumber\\
V_3^\mu &=& \epsilon^{\mu\alpha\beta\gamma} 
            q_{1\alpha} q_{2\beta} q_{3\gamma}\nonumber\\
V_4^\mu &=& q_1^\mu + q_2^\mu + q_3^\mu \equiv Q^\mu \, .
\end{eqnarray}
The vectors $V_{1,2}^\mu$ correspond to an axial-vector intermediate hadronic 
state ({\it e.g.}~the $a_1$ meson), $V_3^\mu$ to a vector intermediate 
hadronic state ({\it e.g.}~the $\rho^\prime$ meson), 
and $V_4^\mu$ to a pseudoscalar or scalar intermediate 
hadronic state ({\it e.g.}~the $\pi^\prime$ meson). 
Accordingly, the form factors $F_{1,2}$ describe 
axial-vector contributions, $F_3$ vector contributions, and $F_4$ pseudoscalar 
or scalar contributions. As can be seen from Eq.~\ref{eq:vmu}, in the reference 
frame ${\cal S} (x,y,z)$ the hadronic current $h^\mu$ is decomposed 
into  the time-like component $h^0$ induced by pseudoscalar
or scalar contributions, the spatial components $h^1$ and $h^2$ by axial-vector
contributions, and the spatial component $h^3$ by vector contributions.
Table \ref{tab:struc_define} summarizes the connection between the different
$J^P$-states, {\it i.e.}~the different components of the hadronic
current $h^\mu$ in the reference frame ${\cal S} (x,y,z)$, and the structure 
functions $W_X$.

\begin{table}[thb]
\begin{center}
\caption[]{\small The different $J^P$-states and their connection with the sixteen 
structure functions $W_X$.}
\label{tab:struc_define}
\begin{tabular}{ c  c  c   c }
  & $J=0$ & $J^P = 1^+$  & $J^P = 1^-$ \\ 
  & $h^0$ & $h^1$, $h^2$ & $h^3$       \\ \hline
  & & & \\
\raisebox{1.4ex}[-1.4ex]{$J=0$}           & ${\bf W_{SA}}$ &  & \\ 
\raisebox{1.4ex}[-1.4ex]{$h^{\star 0}$}  & & &  \\ \hline
  & & &  \\
\raisebox{1.4ex}[-1.4ex]{$J^P = 1^+$}  & 
               \raisebox{1.4ex}[-1.4ex]{$W_{SB}$, $W_{SC}$}  &
               \raisebox{1.4ex}[-1.4ex]{${\bf W_A}$} & \\
\raisebox{1.4ex}[-1.4ex]{$h^{\star 1} $, $h^{\star 2} $}  &  
                      \raisebox{1.4ex}[-1.4ex]{$W_{SD}$, $W_{SE}$} & 
                      \raisebox{1.4ex}[-1.4ex]{$W_C$, $W_D$, $W_E$} & \\ \hline
 & & &  \\
\raisebox{1.4ex}[-1.4ex]{$J^P = 1^-$}  & $W_{SF}$, $W_{SG}$ &
               \raisebox{1.4ex}[-1.4ex]{$W_F$, $W_G$} & ${\bf W_B}$ \\
               \raisebox{1.4ex}[-1.4ex]{ $h^{\star 3}$}  & & 
               \raisebox{1.4ex}[-1.4ex]{$W_H$, $W_I$}   & \\
\end{tabular}
\end{center}
\end{table}

The vector as well as the pseudoscalar contributions are 
suppressed due to $G$-Parity and PCAC. Nonetheless they can contribute. 
For example a possible reaction for the vector contribution in the charged pion 
mode is the decay chain 
$\tau^\mp\rightarrow\rho^{\prime\mp}\nu\rightarrow\pi^\mp\omega\nu$ 
with the $\omega$ meson decaying electromagnetically to two pions 
$\omega\rightarrow\pi^+\pi^-$.
Ref.~\cite{iso-mirkes} estimates that this decay contributes with $0.4\%$ to the
total $\tau^\mp\rightarrow\pi^\mp\pi^\mp\pi^\pm\nu$ decay rate, in agreement
with the measured branching fraction of $0.6\%$ \cite{argus}. 
In the neutral mode, $\tau^\mp\rightarrow\pi^\mp\pi^0\pi^0\nu$, vector 
contributions can occur via $\eta\pi^0$ mixing: 
$\tau^\mp\rightarrow\pi^\mp\pi^0\eta\nu$ with the $\eta$ meson transforming to 
a pion $\eta\rightarrow 2\gamma\rightarrow \pi^0$. 
The average of the measured branching fraction 
$\tau^\mp\rightarrow\pi^\mp\pi^0\eta\nu$ from CLEO \cite{cleo} and ALEPH 
\cite{aleph} is $(0.17\pm 0.03)\%$. Assuming the $\eta\pi^0$ mixing to be 
of the order of $10^{-2}$, ref.~\cite{iso-mirkes} estimates the vector 
contribution in the neutral mode to be of the order of $10^{-5}$.

States with $J=0$ might occur via the $\pi^\prime$ intermediate state:
$\tau^\mp\rightarrow\pi^{\prime\mp}\nu_\tau$. Asumming that the 
$\pi^\prime$ meson decays subsequently to $\rho\pi$ or $\sigma\pi$,
CLEO \cite{cleo_3pi} obtained the following upper limits:
${\cal B} (\tau\rightarrow\pi^\prime\nu_\tau\rightarrow\rho\pi\nu_\tau 
\rightarrow 3\pi\nu_\tau ) < 1.0 \times 10^{-4}$ and 
${\cal B} (\tau\rightarrow\pi^\prime\nu_\tau\rightarrow\sigma\pi\nu_\tau  
\rightarrow 3\pi\nu_\tau ) < 1.9 \times 10^{-4}$ at the $90\%$ confidence level.

Evidently, the sixteen structure functions $W_X$ are not independent 
from each other. They are constructed from the hadronic current $h^\mu$, 
which is determined by seven real numbers (an overall phase is
redundant). Thus, there must exist nine relations among the sixteen structure 
functions. In general these relations are derived from
\begin{equation}
\label{eq:relation}
H_{\alpha\beta} \cdot H_{\gamma\delta}  \equiv
h^\alpha h^{\star\beta}  \cdot h^\gamma h^{\star\delta} =
h^\alpha h^{\star\delta} \cdot h^\gamma h^{\star\beta} \equiv
H_{\alpha\delta} \cdot H_{\gamma\beta} \mbox{ for all } 
\{\alpha , \beta , \gamma , \delta \} \, .
\end{equation}
Employing these relations enables one to deduce the structure functions 
$W_{SC}$, $W_{SE}$, and $W_{SG}$, despite the fact, as mentioned
above, that the leptonic functions $\bar{L}_{SC}$, $\bar{L}_{SE}$,
and $\bar{L}_{SG}$ vanish after integration over the unobservable
$\tau$ azimuth angle $\alpha$.
    

\section{DATA SAMPLE AND EVENT SELECTION}
\label{s-evtsel}

The analysis described here is based on 4 fb$^{-1}$ of $e^+e^-$ 
collision data collected at center-of-mass energies  
$2E_{\rm beam}$ of $\sim 10.6$ GeV, 
corresponding to $4 \times 10^{6}$ reactions of the type 
$e^+e^-\to \tau^+\tau^-$.  These data were recorded at the 
Cornell Electron Storage Ring (CESR) 
with the CLEO~II detector~\cite{cleonim} between 1990 and 1995.  

\subsection{The CLEO~II Detector}

CLEO~II is a general-purpose large solid angle magnetic spectrometer 
and calorimeter.  
Charged particle tracking is accomplished with three concentric 
cylindrical devices: a six-layer straw tube array
surrounding a beam pipe of radius 3.2~cm that encloses 
the $e^+e^-$ interaction point (IP), 
followed by two co-axial drift chambers of 10 and 51 
sense wire layers respectively.  
Barrel ($|\cos\theta|<0.81$, where $\theta$ is the polar 
angle relative to the beam axis) and end cap scintillation counters 
used for triggering and time-of-flight measurements 
surround the tracking chambers.  The calorimeter comprises  
7800 CsI(Tl) crystals, arrayed in projective 
(toward the IP) and axial geometries in barrel and end cap sections 
respectively.  The barrel crystals present 16 radiation lengths to 
photons originating from the IP.  

Identification of $\tau^\mp\to \pi^\mp\pi^0\pi^0\nu_\tau$ decays relies 
heavily on the segmentation and energy resolution of the calorimeter for 
reconstruction of the $\pi^0$'s.  The central portion of the 
barrel calorimeter ($|\cos\theta|<0.71$) achieves energy and 
angular resolutions of 
$\sigma_E/E\,(\%) = 0.35/E^{0.75} + 1.9 - 0.1\,E$ and 
$\sigma_\phi\,\mbox{(mrad)} = 2.8/\sqrt{E} + 1.9
$, with $E$ in GeV, 
for electromagnetic showers.  The angular resolution ensures that 
the two clusters of energy deposited by the photons from a $\pi^0$ 
decay are resolved over the range of $\pi^0$ energies typical of 
the $\tau$ decay mode studied here.

The detector elements described above are immersed in a 1.5 Tesla 
magnetic field provided by a superconducting solenoid surrounding the 
calorimeter.  Muon identification is accomplished with proportional 
tubes embedded in the flux return steel at depths representing 
3, 5 and 7 interaction lengths of total material penetration 
at normal incidence. 

\subsection{Identification of Candidate 
            $\tau^\mp\to \pi^\mp\pi^0\pi^0\nu_\tau$ Decays
           }

The event selection is nearly identical to that used in our 
model dependent analysis~\cite{cleo_3pi}  
of the $\tau^\mp\to\pi^\mp\pi^0\pi^0\nu_\tau$ decay.  
Here, we summarize the main features of the event selection procedure.  
For additional details, the article describing the model dependent 
analysis~\cite{cleo_3pi} should be consulted.  

To identify events as $\tau\tau$ candidates we require the decay of 
the $\tau^\pm$ (denoted as the `tagging' decay) that is 
recoiling against our signal $\tau^\mp$ decay to be classified 
as $\overline{\nu}_\tau e^\pm\nu_e$, or $\overline{\nu}_\tau \mu^\pm\nu_\mu$.
A track is identified as an electron if its calorimeter 
energy to track momentum ratio satisfies $0.85<E/p<1.1$ and if its 
specific ionization in the main drift chamber is not less than $2\sigma$ 
below the value expected for electrons.  
It is classified as a muon if the track has penetrated to at least the 
innermost layer of muon chambers at 3 interaction lengths.  
Thus, we select events that contain two 
oppositely charged barrel tracks ($\vert \cos\theta \vert < 0.81 $)
with momenta between $0.08\, E_{\rm beam}$ and $0.90\, E_{\rm beam}$ and   
separated in angle by at least $90^\circ$, of which one track must be 
identified as an electron or muon.  

Clusters of energy deposition  
in the central region of the calorimeter ($|\cos\theta|<0.71$) 
that are not matched with a charged track projection  
are paired to form $\pi^0$ candidates.  
These showers must have energies greater than 50~MeV, and the 
invariant mass of the photon-pair must lie within $7.5\sigma$ of the 
$\pi^0$ mass where $\sigma$ varies between $\sim 4-7$ MeV.
Those $\pi^0$ candidates with energy above $0.06\, E_{\rm beam}$ 
after application of a $\pi^0$ mass constraint are associated 
with any track within $90^\circ$.  

A $\pi^\mp\pi^0\pi^0$ candidate is formed from a track which has two 
associated $\pi^0$ candidates as defined above.  
If more than one combination of $\pi^0$ candidates can be assigned 
to a given track, only one combination is chosen: namely, that for 
which the largest unused barrel photon-like cluster in the $\pi^\mp\pi^0\pi^0$ 
hemisphere has the least energy.  A cluster is defined to be photon-like 
if it satisfies a $1\%$ confidence level cut on the transverse shower 
profile  
and lies at least 30~cm away from 
the nearest track projection.  

To ensure that these classifications are consistent with 
expectations from $\tau$ decay, events are vetoed if any unused 
photon-like cluster with $|\cos\theta| < 0.95$ 
has energy greater than 200 MeV, or if any 
unmatched non-photon-like cluster has energy above 500 MeV.  
The missing momentum as determined using the $\pi^\mp\pi^0\pi^0$ and tagging 
systems must point into a high-acceptance region of the detector 
($|\cos\theta_{\rm miss}| < 0.9$), and must have a component 
transverse to the beam of at least $0.06\, E_{\rm beam}$.

Finally, 
we define the $\pi^0\pi^0$ signal region to be that where 
the normalized invariant masses of the two photon-pairs,  
$S_{\gamma\gamma} \equiv (M_{\gamma\gamma}-m_{\pi^0})/\sigma_{\gamma\gamma}$,  
satisfies $-3.0 < S_{\gamma\gamma}< 2.0$ for both $\pi^0$ candidates.  
To estimate the contributions from fake $\pi^0$'s, we also define side 
and corner band regions using 
$-7.5< S_{\gamma\gamma} < -5.0$ and $3.0 < S_{\gamma\gamma} < 5.5$. 
The final sample consists of 15849 events in the $\pi^0\pi^0$ signal 
region.  The side and corner band regions contain 1667 and 296 
events, respectively.  

The 
dominant backgrounds are due to mis-identification of $\tau$ decays 
to other final states.  These background modes include (1) decays 
such as $\tau^\mp\to\pi^\mp\pi^0\nu_\tau$ in which a spurious $\pi^0$ 
is reconstructed primarily
from secondary clusters arising from interaction of 
the charged pion in the calorimeter, (2) decay modes with three or 
more $\pi^0$'s, in which the photons associated with the extra $\pi^0$ 
are not detected, (3) the Cabibbo-supressed decay 
$\tau^\mp\to K^\mp\pi^0\pi^0\nu_\tau$, and (4) the mode 
$\tau^\mp\to \pi^\mp K^0_S \nu_\tau$, in which the $\pi^0$'s originate via 
the $K_S\to\pi^0\pi^0$ decay.  The level of 
contamination from these background sources is given in Table \ref{tab:back}.
  
\begin{table}
\begin{center}
\caption[]{\small The four dominant background sources. The total
background contribution is $12.6\%$ leaving an additional $0.2\%$
background contribution due to multiple sources not listed here.}
\label{tab:back}
\begin{tabular}{cccc}
fake $\pi^0$&$\tau^\mp\rightarrow \pi^\mp 3\pi^0 \nu_\tau$& 
$\tau^\mp\rightarrow K^\mp\pi^0\pi^0 \nu_\tau$&$\tau^\mp\rightarrow K_S\pi^\mp \nu_\tau$
\\ \hline
$(8.3 \pm 0.2)\%$&$(3.2 \pm 0.2)\%$&$(0.5 \pm 0.1)\%$&$(0.4 \pm 0.1)\% $ \\ 
\end{tabular}
\end{center}
\end{table}

\section{METHOD}
\label{s-method}

In our model dependent analysis \cite{cleo_3pi}, where we were 
only interested in the relative contributions of the 
amplitudes considered, we used a single entry maximum likelihood fit. 
Here, we extend the single entry maximum likelihood by the 
normalization to measure the absolute rate in a given bin of 
$Q^2$ as well as $(Q^2 ,s_1 , s_2 )$. As mentioned in section 
\ref{s-intro} these rates are determined by the structure 
functions $W_A$, $W_B$, and $W_{SA}$.

The likelihood in a given bin $j$ of $Q^2$ or $(Q^2 , s_1 , s_2 )$ 
can be written as follows
\begin{equation}
\label{eq:likel}
-2 \ln {\cal L}_j = \sum_{i}^{N_{evt, j}} 
\biggl[  -2 \ln \sum_{X}  
{ \tilde{L}_X W_X (Q^2 , s_1 , s_2  ) \brace \tilde{L}_X w_X (Q^2) }  
\biggr] + 2 N_j \, ,
\end{equation}
where $N_{evt,j}$ is the number of events in bin $j$,
and $N_j$ the normalization of bin $j$. The upper expression in the 
braces in Eq.~\ref{eq:likel} is used in the determination of the 
structure functions $W_X$ differential in the Dalitz plot, and the lower
expression for the determination of the structure functions $w_X$
integrated over the Dalitz plot.
The leptonic functions $\tilde{L}_X$ are integrated over the unobservable 
$\tau$ lepton azimuthal angle $\alpha$ including the $\tau$ pair 
production $P$ and the factorized initial state radiation $f_{ini}$
\begin{equation}
\label{eq:ltilde}
\tilde{L}_X (\beta , \gamma , \theta_\tau , Q^2 )
= \int f_{ini} (\vec{\kappa})\cdot  P\cdot 
\bar{L}_X (\alpha , \beta , \gamma , \theta_\tau , Q^2 ) d\vec{\kappa}
d\alpha \, ,
\end{equation}
where $\vec{\kappa}$ is the momentum of the radiated photon. 
For the initial state photon spectrum $f_{ini}$ we use the 
formula obtained in ref.~\cite{behrends}.
Overall constant factors of the $\tau$ pair production have 
been factored out and, accordingly, $P$ denotes only the functional form 
of the $\tau$ pair production with $P = (1 + \cos^2\theta_{\tau\tau} )/2$, 
where $\theta_{\tau\tau} $ is the polar angle of the $\tau$ lepton 
with respect to the beam axis. The integrals over $\alpha$ and $\vec{\kappa}$ 
in Eq.~\ref{eq:ltilde} are evaluated numerically.

We have included the initial state bremsstrahlung in the leptonic 
functions $\tilde{L}_X$ to account for the fact that
the $\tau$ energy $E_\tau$ in the laboratory frame, needed for the evaluation 
of the angle $\theta_\tau$ (Eq.~\ref{eq:dgamma}), is in general not given  
by $E_\tau = \sqrt{s}/2$ but by $E_\tau = \sqrt{s^\prime} /2$, 
where $\sqrt{s^\prime}$ is the center of mass energy after radiation. 

The inclusion of the $\tau$ pair production corrects for the fact that 
the $(1 + cos^2\theta_{\tau\tau})$ distribution of the $\tau$ leptons induces an
anisotropic distribution of the $\tau$ momenta that lie on a cone around the 
three pion momentum. Azimuth angles that result in $\tau$ momenta closer to the
beam are preferred over azimuth angles further away from the beam. In the
absence of any selection cuts this anisotropy vanishes after integration over the
orientation of the $\tau$ lepton or, equivalently, the orientation of the three 
pion system in the laboratory frame. However, the selection cuts applied yield
a residual anisotropy affecting the leptonic functions $\bar{L}_X$.

Monte Carlo studies showed that the neglect of the $\tau$ pair production as well 
as of the initial state radiation results in a bias in the determination of the
structure functions. In both cases the bias obtained is around $2\%$. 
Although the statistical uncertainty in our results is much larger than the 
effect of $\tau$ pair production, we expect that future analyses will not be
limited by statistics and that pair production dynamics will be more important.

For a given bin $j$ in $Q^2$ or in $(Q^2, s_1 , s_2 )$ the normalization $N_j$ in 
Eq.~\ref{eq:likel} is chosen to be 
\begin{equation}
\label{eq:norm}
N_j = \frac{ N_{evt} }{ f_{sel} \Gamma } \times  \tilde{\Gamma}_j 
    = \frac{ N_{evt} }{ f_{sel} \Gamma } \times  
C ( Q^2 )   \sum_{X}  \int 
{ \tilde{L}_X W_X (Q^2 , s_1 , s_2 ) \brace \tilde{L}_X w_X (Q^2) } 
d\gamma d\cos\beta d\cos\theta_\tau \, ,
\end{equation}
where $\beta$, $\gamma$, $\theta_\tau$, and $C( Q^2 )$ have been 
introduced above (see Eq.~\ref{eq:dgamma}). 
The total number of events is denoted by  $N_{evt}$.
The partial width, $\Gamma$, of the $\tau^\mp$ lepton decaying into 
$\pi^\mp\pi^0\pi^0\nu_\tau$ is given by the total width 
$\Gamma_{tot}  = \hbar / \tau_\tau$ times the branching fraction 
${\cal B} ( \tau^\mp\rightarrow\pi^\mp\pi^0 \pi^0 \nu_\tau )$. 
We use the world average values of 
$\Gamma_{tot} = (2.27 \pm 0.01 ) \times 10^{-3} \mbox{ eV}$ 
\cite{pdg} and 
${\cal B} ( \tau^\mp\rightarrow\pi^\mp\pi^0 \pi^0 \nu_\tau ) = (9.15 \pm 0.15)\%$ 
\cite{pdg} to determine the partial width $\Gamma$.
The factor $f_{sel} = 0.996$ corrects for events that 
are outside of our selected region of 
$0.5 \mbox{ GeV}^2/\mbox{c}^4 < Q^2 < 2.75  \mbox{ GeV}^2/\mbox{c}^4$, 
{\it i.e.}~$\sum_j \tilde{\Gamma}_j = f_{sel} \times \Gamma$.

As in our model dependent anaylysis \cite{cleo_3pi}, we account for 
the four main background sources, listed in Table \ref{tab:back}, by 
extending the likelihood  of Eq.~\ref{eq:likel} as follows
\begin{eqnarray}
{\cal L} & = &
( 1 - \alpha_{f_{\pi^0}} - \alpha_{4\pi} 
    - \alpha_{K\pi\pi} -\alpha_{K_S \pi} ) 
                   {\cal L}_{signal}  \nonumber \\
& & \mbox{} +
\alpha_{f_{\pi^0}} {\cal L}_{f_{\pi^0}} +
\alpha_{4\pi}      {\cal L}_{4\pi}     +
\alpha_{K\pi\pi}   {\cal L}_{K\pi\pi} +
\alpha_{K_S \pi}   {\cal L}_{K_S \pi} \, ,
\end{eqnarray}
where ${\cal L}_{signal}$ is the likelihood of the signal events,
${\cal L}_{f_{\pi^0}}$ the likelihood of the fake $\pi^0$ background,
${\cal L}_{4\pi}$ the likelihood of the 
$\tau^\mp\rightarrow\pi^\mp ( 3\pi^0 ) \nu_\tau$ background, 
${\cal L}_{K\pi\pi}$ the likelihood of the 
$\tau^\mp\rightarrow K^\mp \pi^0\pi^0\nu_\tau$ background, and
${\cal L}_{K_S \pi}$ the likelihood of the 
$\tau^\mp\rightarrow K_S \pi^\mp \nu_\tau$ background.
The corresponding background fractions $\alpha_i$ are given by 
Table \ref{tab:back}. The likelihood ${\cal L}_{f_{\pi^0}}$ of the fake pion
background is approximated by the Dalitz plot distributions of events
populating the $\pi^0$ mass side bands. We model the likelihood 
${\cal L}_{4\pi}$ of the four pion background with the decay 
$\rho (1450) \rightarrow \rho \sigma$ using a $S$-wave amplitude. To estimate
the systematic error arsising from the large theoretical and experimental 
uncertainites of the four pion matrix element we also consider the decay 
$\rho (1450) \rightarrow a_1 \pi \mbox{ ($S$-wave)}$. 
The integration over the lost $\pi^0$ meson, needed for the evaluation of 
${\cal L}_{4\pi}$, is done numerically taking into account efficiency.
The background $\tau^\mp\rightarrow  K^\mp\pi^0\pi^0\nu_\tau $ is modeled
by the decay chain
$ \tau^\mp\rightarrow  K_1^\mp \nu_\tau \mbox{, }  
K_1^\mp \rightarrow K^{\star \mp} \pi^0 \mbox{ ($S$-wave), }$ 
where the $K_1$ meson is parametrized by
a superposition of the $K_1 (1270)$ and $K_1(1400)$ Breit Wigner functions.  
Finally, the $\tau^\mp\rightarrow K_S\pi^\mp \nu_\tau $ background is 
parametrized by the decay chain $\tau^\mp \rightarrow  K^{\star \mp} \nu_\tau$, 
$K^{\star \mp} \rightarrow K_S^0 \pi^\mp$ ($P$-wave). 
The mass distribution for the $K_S \rightarrow \pi^0\pi^0$ decay 
is parametrized by a Gaussian, where the mean and the width are taken from data.

To determine the structure function integrated over the Dalitz plot
we subdivide our data sample in nine equidistant bins of $Q^2$. The 
considered $Q^2$ range is 
$0.50\mbox{ GeV}^2/\mbox{c}^4 < Q^2 < 2.75 \mbox{ GeV}^2/\mbox{c}^4$
yielding a bin width of $0.25\mbox{ GeV}^2/\mbox{c}^4$. In each of the nine
bins we fit for the integrated structure functions $w_A$, $w_C$, $w_D$, 
and $w_E$. In this case non-axial-vector contributions are not taken 
into consideration, since, due to the loss of information by integrating
over the Dalitz plot, we are unable to resolve these contributions. 

Table \ref{tab:bin_def} shows in detail the  binning in $(Q^2 , s_1 , s_2 )$ 
used to extract the structure functions $W_X$ differential in the Dalitz plot.
In total we have 34 bins. Instead of fitting directly for the structure
functions $W_X$ and employing explicitly the nine relations among the structure
functions (see Eq.~\ref{eq:relation}), we introduce in each bin of
$(Q^2 , s_1 , s_2 )$ seven fit parameters:
the real $\Re( h_0 )$ and imaginary part $\Im( h_0 )$ of the time-like 
component of the hadronic current $h^\mu$, the real part $\Re (h_1 )$
of the $x$-component, the real $\Re (h_2 )$ and imaginary part $\Im (h_2 )$
of the $y$-component, and the real $\Re (h_3 )$ and imaginary part $\Im (h_3 )$
of the $z$-component. The imaginary part of the $x$-component $\Im (h_1) $
is chosen to be zero. From the measured hadronic current $h^\mu$ in a given
bin of $(Q^2 , s_1 , s_2 )$ we then calculate the structure functions
$W_X$ taking into account the full covariance matrix for the error evaluation
of the structure functions $W_X$.

\begin{table}[thb]
\begin{center}
\caption[]{\small Employed binning in $(Q^2 , s_1 , s_2 )$ as used in the 
fits differential in the Dalitz plot. The Dalitz plane is symmetrized 
with $s_1 > s_2$. Depending on $s_1$, $s_2^{max}$ is either given by the 
phase space boundaries (see {\it e.g.}~ref.~\cite{pdg}, section $35.4.3.1$)
or by the condition $s_1 > s_2$.}
\label{tab:bin_def}
\renewcommand{\arraystretch}{0.9}
{\small
\begin{tabular}{ r  c c c  r }
Bin&$Q^2\mbox{ }[\mbox{GeV}^2 ]$&$s_1\mbox{ }[\mbox{GeV}^2]$&
$s_2\mbox{ }[\mbox{GeV}^2]$&$N_j$\\
\hline
 1&                                     &$0.00-0.35$&$0.00-s_2^{max}$&$192$\\
 2&\raisebox{1.4ex}[-1.4ex]{$0.50-0.75$}&$0.35-0.75$&$0.00-s_2^{max}$&$204$\\
\hline
 3&                                     &$0.00-0.48$&$0.00-0.32     $&$527$\\
 4&                                     &$0.48-0.56$&$0.00-0.32     $&$575$\\
 5&\raisebox{1.4ex}[-1.4ex]{$0.75-1.00$}&$0.56-1.00$&$0.00-0.32     $&$492$\\
 6&                                     &$0.32-0.68$&$0.32-s_2^{max}$&$512$\\
\hline
 7&                                     &$0.00-0.53$&$0.00-0.40     $&$680$\\
 8&                                     &$0.53-0.59$&$0.00-0.40     $&$619$\\
 9&                                     &$0.59-0.66$&$0.00-0.40     $&$663$\\
10&\raisebox{1.4ex}[-1.4ex]{$1.00-1.25$}&$0.66-1.25$&$0.00-0.40     $&$614$\\
11&                                     &$0.40-0.56$&$0.40-s_2^{max}$&$645$\\
12&                                     &$0.56-0.85$&$0.40-s_2^{max}$&$667$\\
\hline
13&                                     &$0.00-0.56$&$0.00-0.52     $&$667$\\ 
14&                                     &$0.56-0.64$&$0.00-0.52     $&$675$\\
15&                                     &$0.64-0.74$&$0.00-0.52     $&$643$\\
16&\raisebox{1.4ex}[-1.4ex]{$1.25-1.50$}&$0.74-1.50$&$0.00-0.52     $&$663$\\
17&                                     &$0.52-0.66$&$0.52-s_2^{max}$&$665$\\
18&                                     &$0.66-0.98$&$0.52-s_2^{max}$&$640$\\
\hline
19&                                     &$0.00-0.61$&$0.00-0.57     $&$496$\\
20&                                     &$0.61-0.73$&$0.00-0.57     $&$481$\\ 
21&                                     &$0.73-0.89$&$0.00-0.57     $&$442$\\
22&\raisebox{1.4ex}[-1.4ex]{$1.50-1.75$}&$0.89-1.75$&$0.00-0.57     $&$440$\\
23&                                     &$0.57-0.77$&$0.57-s_2^{max}$&$490$\\
24&                                     &$0.77-1.18$&$0.57-s_2^{max}$&$498$\\    
\hline
25&                                     &$0.00-0.72$&$0.00-0.66     $&$382$\\
26&                                     &$0.72-1.00$&$0.00-0.66     $&$379$\\
27&\raisebox{1.4ex}[-1.4ex]{$1.75-2.00$}&$1.00-2.00$&$0.00-0.66     $&$360$\\ 
28&                                     &$0.66-1.34$&$0.66-s_2^{max}$&$373$\\
\hline
29&                                     &$0.00-1.05$&$0.00-s_2^{max}$&$320$\\
30&\raisebox{1.4ex}[-1.4ex]{$2.00-2.25$}&$1.05-2.25$&$0.00-s_2^{max}$&$300$\\  
\hline
31&                                     &$0.00-1.25$&$0.00-s_2^{max}$&$109$\\
32&\raisebox{1.4ex}[-1.4ex]{$2.25-2.50$}&$1.25-2.50$&$0.00-s_2^{max}$&$124$\\ 
\hline
33&                                     &$0.00-1.20$&$0.00-s_2^{max}$&$55$\\ 
34&\raisebox{1.4ex}[-1.4ex]{$2.50-2.75$}&$1.20-2.75$&$0.00-s_2^{max}$&$47$\\
\end{tabular}
}
\end{center}
\end{table}

\section{MEASUREMENT}
\label{s-measure}

\begin{figure}[thb]
\begin{center}
\leavevmode
\unitlength1.0cm
\epsfxsize=14.cm
\epsfysize=12.cm
\epsffile{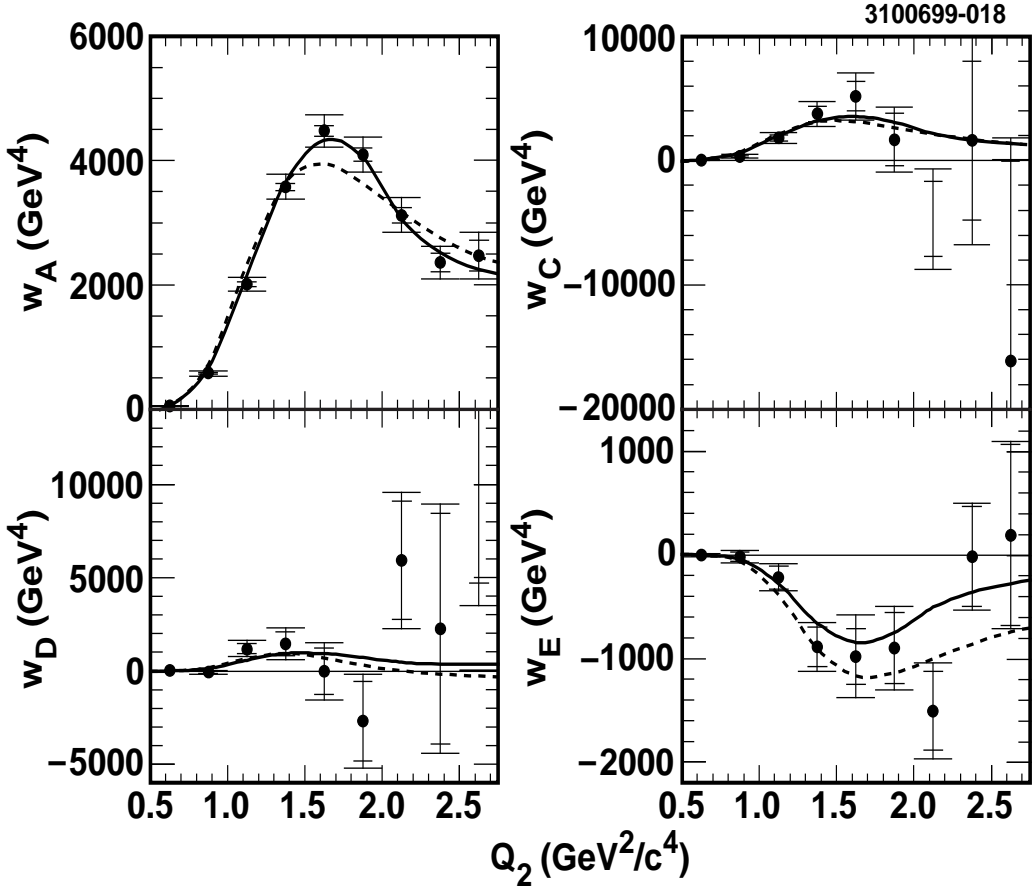}
\caption[]{\small The integrated structure functions $w_A$, $w_C$, 
$w_D$, and $w_E$. 
The filled points represent the data. The smaller error bars indicate
the statistical errors. The distance between the smaller and larger error
bars shows the systematic errors (see Section \ref{s-sys}). The solid line 
is the model as obtained in our model dependent analysis, whereas the dotted 
line corresponds to the
KS-model.} 
\label{wa_c_d_e}
\end{center}
\end{figure}

The fit results obtained for the integrated structure functions $w_A$,
$w_C$, $w_D$, and $w_E$  in comparison with the model obtained in our 
model dependent analysis \cite{cleo_3pi} as well as the model of
J.~H.~K\"uhn and A.~Santamaria \cite{kuhn-sant} (KS) are shown in 
Fig.~\ref{wa_c_d_e}. The observed steep falling of the structure
function $w_A$ around $Q^2 = 2 \mbox{ GeV}^2/\mbox{c}^4$, which might
be an indication of the $K^\star K$ threshold opening, is well 
reproduced by the fit obtained in our model dependent analysis
\cite{cleo_3pi}. 
The KS-model \cite{kuhn-sant}, which does not include the $K^\star K$ 
threshold in the total width of the $a_1$ meson Breit Wigner function, 
falls less steeply around $Q^2 = 2 \mbox{ GeV}^2/\mbox{c}^4$, 
as can be seen from Fig.~\ref{wa_c_d_e}.
Owing to the large statistical errors obtained on the structure
functions $w_C$, $w_D$, and $w_E$ the comparison of the fit 
results and the models is not very conclusive for those three
functions.

\begin{figure}[thb]
\begin{center}
\leavevmode
\unitlength1.0cm
\epsfxsize=14.cm
\epsfysize=12.cm
\epsffile{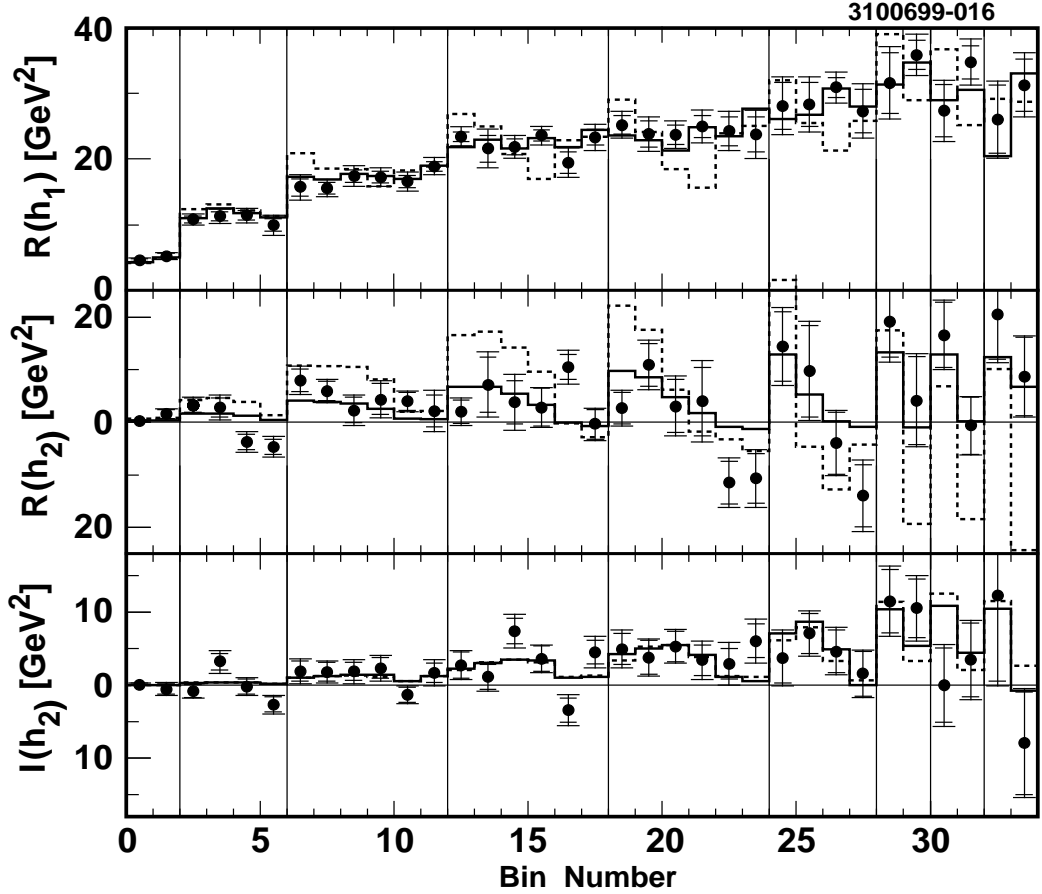}
\caption[]{\small The real and imaginary parts of the axial-vector 
induced components of the hadronic current as measured in our fits 
differential in the Dalitz plot. For the definition of the bin 
number the reader is referred to Table \ref{tab:bin_def}. 
The filled points represent the data. The smaller error bars indicate
the statistical errors. The distance between the smaller and 
larger error bars shows the systematic errors (see Section \ref{s-sys}). 
The solid line is the model as obtained in our model dependent 
analysis, whereas the dotted line corresponds to the KS-model.} 
\label{h_ax_s1s2}
\end{center}
\end{figure}

\begin{figure}[thb]
\begin{center}
\leavevmode
\unitlength1.0cm
\epsfxsize=14.cm
\epsfysize=14.cm
\epsffile{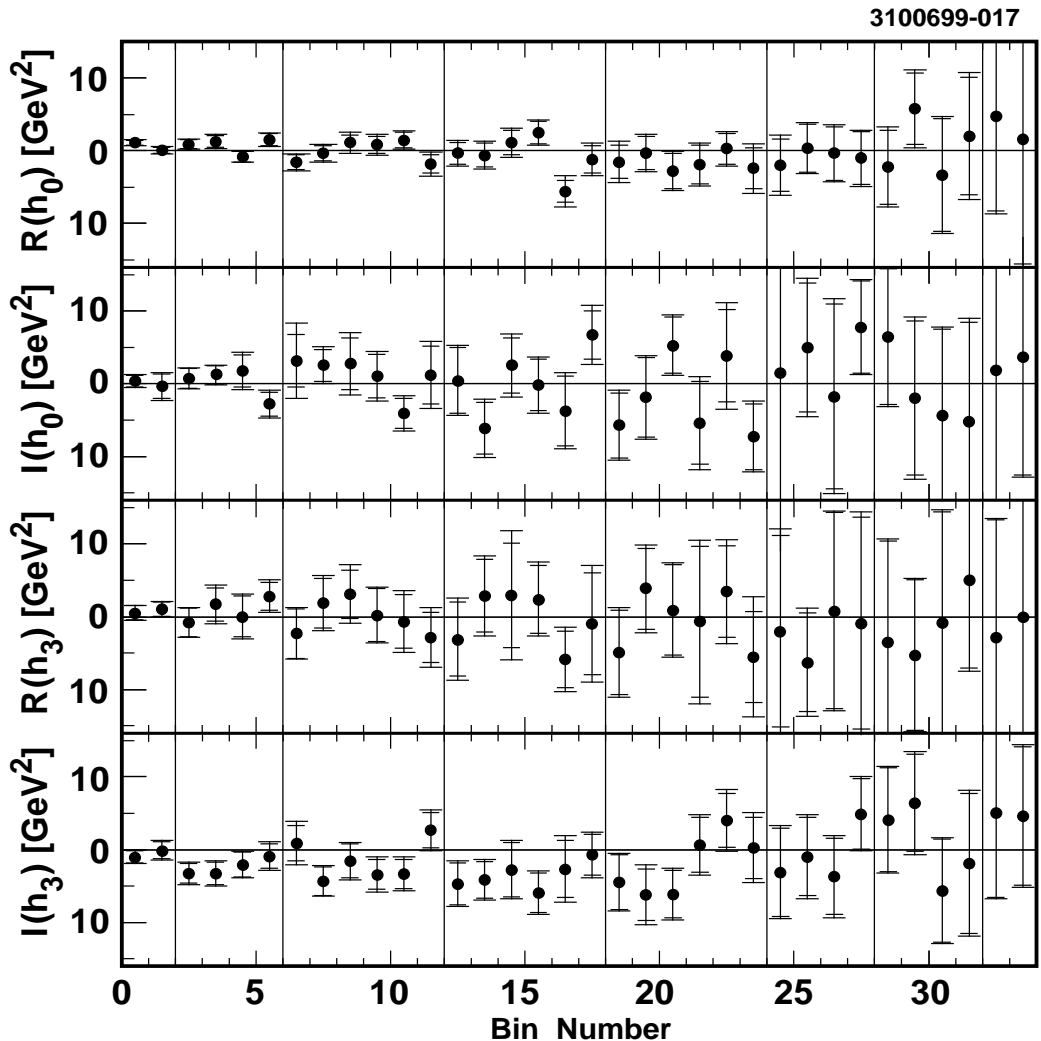}
\caption[]{\small The real and imaginary parts of the non-axial-vector
induced components of the hadronic current as measured in our fits 
differential in the Dalitz plot. 
For further explanations see comments in caption for Fig.~\ref{h_ax_s1s2}.}
\label{h_no_ax_s1s2}
\end{center}
\end{figure}

The fit results for the axial-vector part of the hadronic current 
(Table \ref{tab:struc_define}) in bins of $(Q^2 , s_1 , s_2)$ are shown 
in Fig.~\ref{h_ax_s1s2}.
The measured non-axial-vector contributions of the hadronic current 
are plotted in Fig.~\ref{h_no_ax_s1s2}. 

\begin{figure}[thb]
\begin{center}
\leavevmode
\unitlength1.0cm
\epsfxsize=14.cm
\epsfysize=14.cm
\epsffile{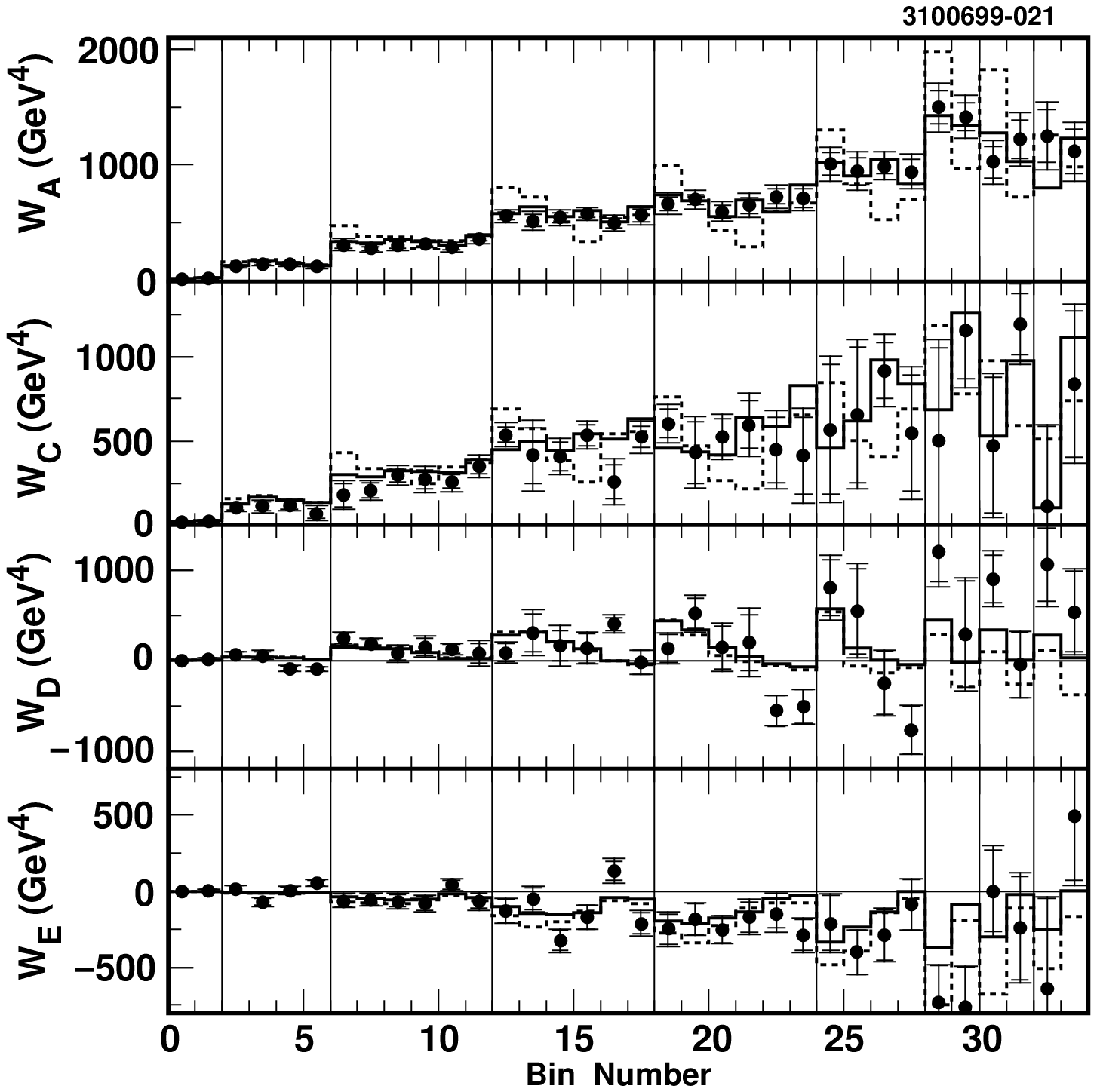}
\caption[]{\small The axial-vector induced structure functions 
$W_A$, $W_C$, $W_D$, and $W_E$ as obtained from our fits to the 
real and imaginary parts of the hadronic current. 
For further explanations see comments in caption for Fig.~\ref{h_ax_s1s2}.}
\label{w_ax_s1s2}
\end{center}
\end{figure}

\begin{figure}[thb]
\begin{center}
\leavevmode
\unitlength1.0cm
\epsfxsize=14.cm
\epsfysize=14.cm
\epsffile{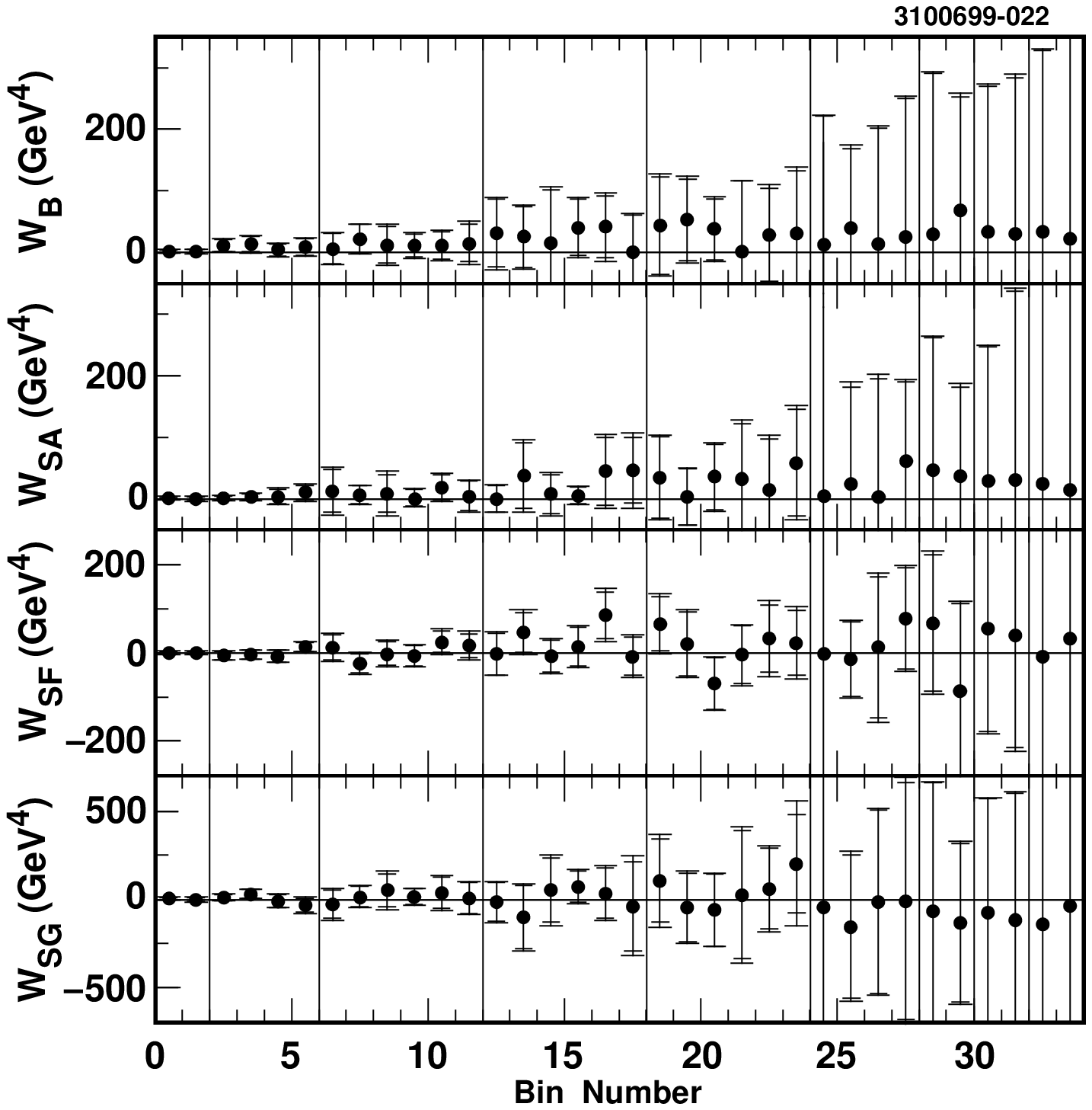}
\caption[]{\small The non-axial-vector induced structure functions 
$W_B$, $W_{SA}$, $W_{SF}$, and $W_{SG}$ as obtained from our fits 
to the real and imaginary parts of the hadronic current.
For further explanations see comments in caption for Fig.~\ref{h_ax_s1s2}.} 
\label{w_vs_s1s2}
\end{center}
\end{figure}

\begin{figure}[thb]
\begin{center}
\leavevmode
\unitlength1.0cm
\epsfxsize=14.cm
\epsfysize=14.cm
\epsffile{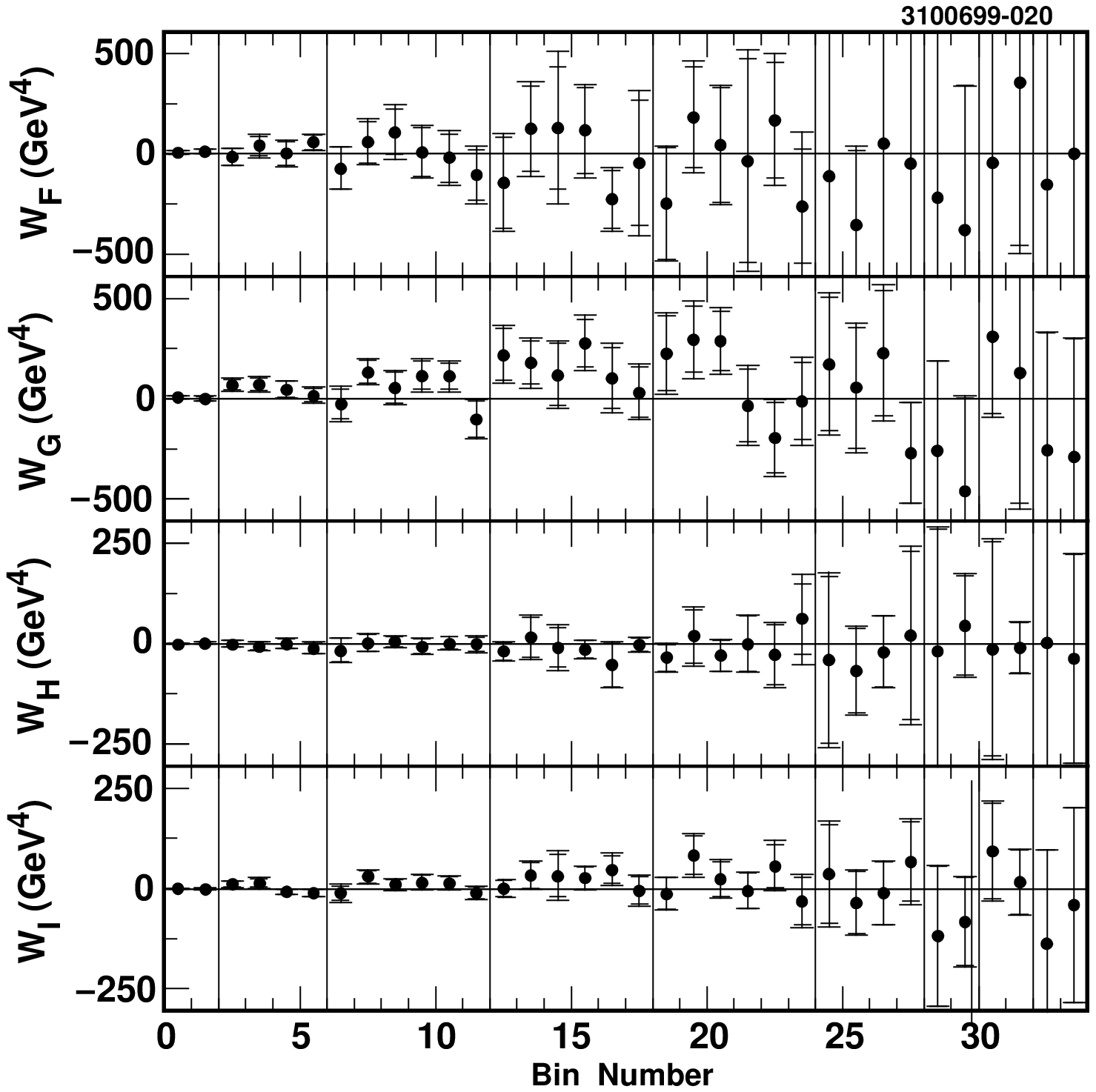}
\caption[]{\small The structure functions $W_F$, $W_G$, $W_H$, 
and $W_I$ --- induced by the interference between the vector and 
axial-vector components  of the hadronic current ---
as obtained from our fits to the real and imaginary parts of 
the hadronic current.
For further explanations see comments in caption for Fig.~\ref{h_ax_s1s2}.}  
\label{w_av_s1s2}
\end{center}
\end{figure}

\begin{figure}[thb]
\begin{center}
\leavevmode
\unitlength1.0cm
\epsfxsize=14.cm
\epsfysize=14.cm
\epsffile{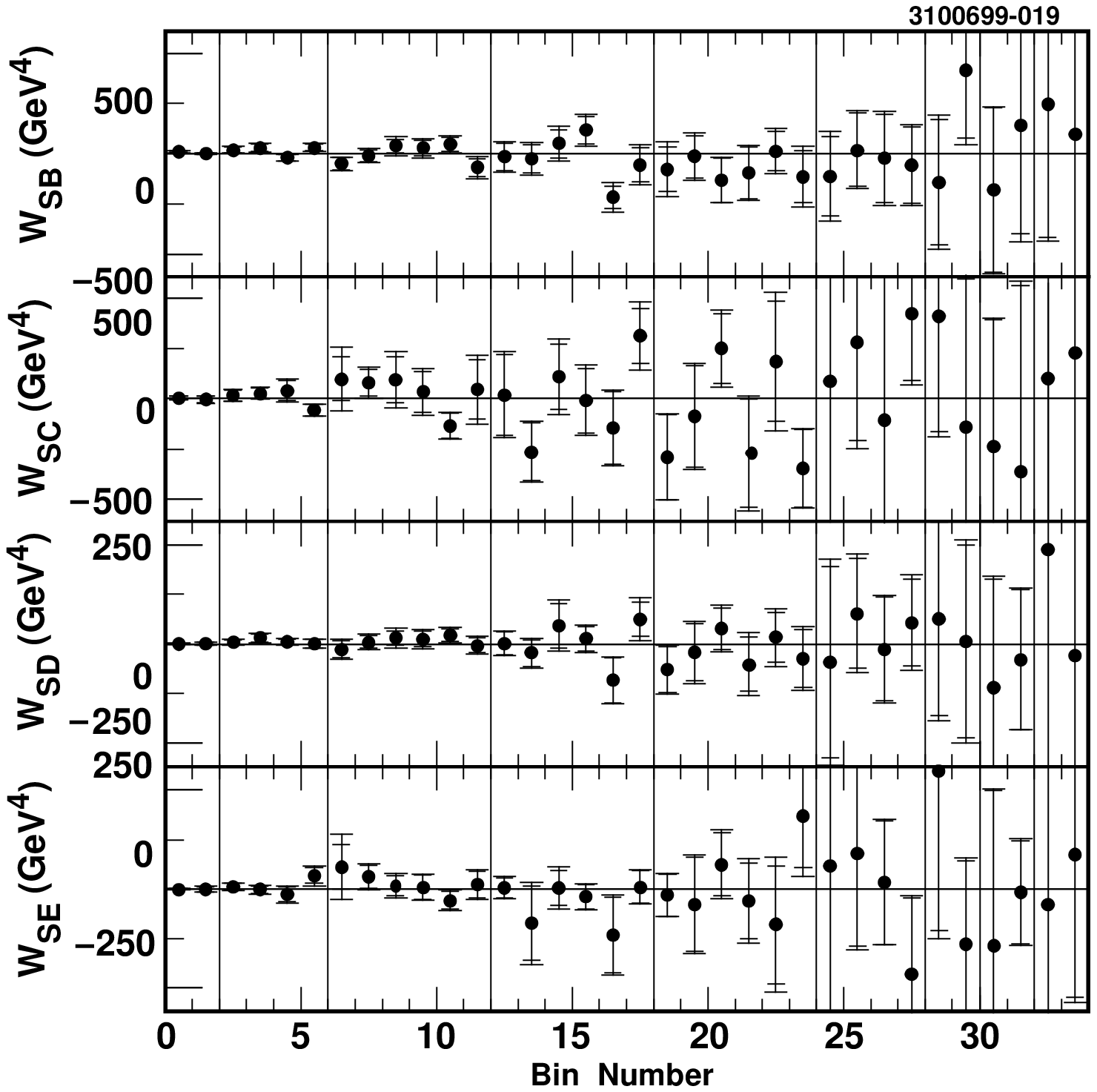}
\caption[]{\small The structure functions $W_{SB}$, $W_{SC}$, $W_{SD}$,
and $W_{SE}$ --- induced by the interference between the scalar and
axial-vector components of the hadronic current --- 
as obtained from our fits to the real and imaginary parts of 
the hadronic current. 
For further explanations see comments in caption for Fig.~\ref{h_ax_s1s2}.}
\label{w_as_s1s2}
\end{center}
\end{figure}

As mentioned in section \ref{s-method}, we derive the structure function
$W_X$ differential in the Dalitz plot from our measured hadronic current
including the full covariance matrix. 
The results obtained on the axial-vector induced structure functions 
$W_A$, $W_C$, $W_D$, and $W_E$ are shown in Fig.~\ref{w_ax_s1s2}, whereas
Fig.~\ref{w_vs_s1s2} shows the non-axial-vector structure functions
$W_{B}$, $W_{SA}$, $W_{SF}$, and $W_{SG}$. The structure functions that have
their origin in the interference between the vector and axial-vector 
contributions of the hadronic current as well as the ones due to the
interference between the scalar and axial-vector components of the 
hadronic current are plotted in Figs.~\ref{w_av_s1s2} and \ref{w_as_s1s2},
respectively.

The KS-model \cite{kuhn-sant}, as can be seen from Fig.~\ref{h_ax_s1s2}, 
fails to describe the details of the hadronic current components $\Re ( h_1 )$ 
and $\Re ( h_2 )$. This discrepancy between the data and the KS-model
\cite{kuhn-sant} is also reflected in the functions $W_A$ and $W_C$, 
shown in Fig.~\ref{w_ax_s1s2}.

As illustrated by Fig.~\ref{h_no_ax_s1s2} and, correspondingly, by the
Figs.~\ref{w_vs_s1s2}, \ref{w_av_s1s2}, and \ref{w_as_s1s2}, we do not
measure significant scalar or vector contributions.
The null hypotheses for scalar and vector contributions as well
as for scalar and vector contribution combined yield in all three
cases significances below three standard deviations.

Tables summarizing our results on the structure functions as well
as on the hadronic current components are available 
on the 
WWW\footnote{See {\bf http://www.lns.cornell.edu/public/CLEO/analysis/results/tau-struct/}.}.

\section{LIMITS ON NON AXIAL-VECTOR CONTRIBUTION}
\label{s-nonaxial}

The fit results on the scalar components of the hadronic current,
{\it i.e.}~the time-like components $\Re (h_0 )$ and  $\Im ( h_0 )$,
yield the following branching fraction for scalar contributions
\begin{equation}
\frac{{\cal B}(\tau^\mp\rightarrow S\nu\rightarrow\pi^\mp\pi^0\pi^0\nu )}
     {{\cal B}(\tau^\mp\rightarrow\pi^\mp\pi^0\pi^0\nu )} =
( 5.2 \pm 2.5 \pm 0.3 )\% \, ,
\end{equation}
where the first error is statistical and the second error systematic.
The individual contributions to the systematic error are:
$\pm 0.29 \%$  due to efficiency, $\pm 0.06\%$ due to Monte Carlo
statistics, $\pm 0.03\%$ due to the energy scale of the photons, 
and $\pm 0.01 \%$ due to background.
With this we obtain the following upper limit
\begin{equation}
\frac{{\cal B}(\tau^\mp\rightarrow S\nu\rightarrow\pi^\mp\pi^0\pi^0\nu )}
     {{\cal B}(\tau^\mp\rightarrow\pi^\mp\pi^0\pi^0\nu )} <
9.4 \% \mbox{ at 95\% CL.}
\end{equation}

Using the  measured contributions of the vector components 
$\Re (h_3 ) $ and $\Im ( h_3 ) $ we measure the branching ratio 
for vector contributions to be
\begin{equation}
\frac{{\cal B}(\tau^\mp\rightarrow V\nu\rightarrow\pi^\mp\pi^0\pi^0\nu )}
     {{\cal B}(\tau^\mp\rightarrow\pi^\mp\pi^0\pi^0\nu )} =
( 4.2 \pm 1.9 \pm 0.2 ) \% \, ,
\end{equation} 
where the first error is statistical and the second error systematic.
The systematic error is given by the quadratic sum of a $\pm 0.15\%$
contribution due to efficiency, a $\pm 0.10\%$ contribution due to Monte Carlo
statistics, a $\pm 0.02\%$ contribution due to the energy scale of the photons,
and a $\pm 0.01\%$ contribution due to background.
The upper limit on vector contributions is
\begin{equation}
\frac{{\cal B}(\tau^\mp\rightarrow V\nu\rightarrow\pi^\mp\pi^0\pi^0\nu )}
     {{\cal B}(\tau^\mp\rightarrow\pi^\mp\pi^0\pi^0\nu )} <
7.3 \% \mbox{ at 95\% CL.}
\end{equation}

Combining the results on the scalar and vector components of the hadronic
current we obtain the following branching ratio for non-axial-vector
contributions
\begin{equation}
\frac{{\cal B}(\tau^\mp\rightarrow \mbox{Non-Axial }\nu\rightarrow\pi^\mp\pi^0\pi^0\nu )}
     {{\cal B}(\tau^\mp\rightarrow\pi^\mp\pi^0\pi^0\nu )} =
( 9.4 \pm 4.4 \pm 0.4 )\% \, ,
\end{equation} 
where the first error is statistical including the correlations between the
vector and scalar contributions and the second error systematic.
The systematic error is given by the quadratic sum of the systematic error of 
$\pm 0.30\%$ from the scalar contribution and the systematic error of 
$\pm 0.18\%$ from the vector contribution. 
The upper limit on non-axial-vector contributions is
\begin{equation}
\frac{{\cal B}(\tau^\mp\rightarrow \mbox{Non-Axial }\nu\rightarrow\pi^\mp\pi^0\pi^0\nu )}
     {{\cal B}(\tau^\mp\rightarrow\pi^\mp\pi^0\pi^0\nu )} <
16.6 \% \mbox{ at 95\% CL.}
\end{equation}

The corresponding upper limit obtained by the OPAL 
collaboration \cite{opal} is
\begin{equation}
\frac{\Gamma^{non-AV} ( \tau^- \rightarrow \pi^- \pi^- \pi^+ \nu )}%
{\Gamma^{tot} ( \tau^- \rightarrow \pi^- \pi^- \pi^+ \nu )} < 26.1\%   
\mbox{ at 95\% CL.}   
\end{equation}

\section{SYSTEMATIC ERRORS}
\label{s-sys}

We consider the following sources of systematic errors: 
\begin{itemize}
\item Overall Normalization: As mentioned in section \ref{s-method}, 
      we normalize our likelihood to the world average branching ratio of 
      ${\cal B} (\tau^\mp\rightarrow\pi^\mp\pi^0\pi^0\nu_\tau) = (9.15 \pm 0.15)\%$.
      The error on the branching ratio results in an overall systematic 
      error of our normalization. In addition, the extrapolation of our 
      measurement into the regions  $Q^2 < 0.5 \mbox{GeV}^2/\mbox{c}^4$ 
      and $Q^2 > 2.75\mbox{GeV}^2/\mbox{c}^4$, {\it i.e.}~the uncertainty 
      on $f_{sel}$ in Eq.~\ref{eq:norm}, also yields an error on our normalization. 
      We conservatively estimated this error to be as large as the error 
      on the branching ratio. Accordingly, we have an overall systematic
      error on all structure functions of $3\%$ and, in case of 
      the hadronic current components, an  overall systematic error of $1.7\%$. 
\item Monte Carlo Statistics: The limited statistics in the calculation of
      the normalization integrals of the likelihood function for the different 
      bins in $Q^2$ as well as in  $(Q^2, s_1 , s_2)$ results in a systematic
      error of our measurement. We estimated this error by subdividing our
      Monte Carlo sample, which has approximately ten times the size of our 
      data sample, in six independent sub-samples. The variation in the
      fit results obtained are used as an estimate for this error.
\item Efficiency: The not fully accurate detector simulation of the
      efficiency of the momenta of the  three pions and the angle between the pions 
      as well as the invariant mass of the three pions can also effect our measurement.
      We estimate this uncertainty by reweighting our events as a function 
      of the momenta of the three pions, the angle between the pions, and the
      invariant mass of the three pions. Since our likelihood is normalized
      to the world average branching ratio \cite{pdg}, the error in the overall 
      efficiency does not affect our measurement. Accordingly, we only account for 
      relative changes of the efficiency.
\item Background: The amount of background as well as the background models  
      used in the likelihood have uncertainties. The systematic error associated 
      with these uncertainties have been estimated by varying in the likelihood
      function the background contribution and, in case of the four pion background, 
      also the model, {\it i.e.}~(1) $\rho (1450 ) \rightarrow \rho \sigma$,
      (2) $\rho (1450 ) \rightarrow a_1 \pi$, and (3) combinations thereof 
      (see section \ref{s-method}).
\item Photon Energy Scale: The energy scale of the photons that
      form the neutral pions has an uncertainty of around $0.3\%$. The resulting
      systematic uncertainty in our measurement has been estimated by 
      rescaling the energies of the neutral pions by $0.3\%$ and refitting our data.      
\end{itemize}

In addition to the systematic uncertainties listed above, we also have an uncertainty 
due to the finite detector resolution. Refitting our events including the error
matrix on the charged track and the photons (scaled by a factor of four) showed that
this source is negligible compared to the other sources and has, therefore, not been
taken into account. This coincides with our findings of our model dependent analysis
\cite{cleo_3pi}.

In general all five sources result in systematic errors that have roughly the
same order of magnitude. However, the errors due to the overall normalization 
and the Monte Carlo statistics yield the largest systematic uncertainy for 
most of the bins.

The systematic errors obtained are shown in Figs.~\ref{wa_c_d_e},
\ref{h_ax_s1s2}, \ref{h_no_ax_s1s2}, \ref{w_ax_s1s2}, 
\ref{w_vs_s1s2}, \ref{w_av_s1s2}, and \ref{w_as_s1s2} as extension
to the error bars.

\section{SUMMARY}
\label{s-summ}

We have measured the integrated structure functions
$w_A$, $w_C$, $w_D$, and $w_E$. In addition, we determine
the seven non-trivial real and imaginary components of the 
hadronic current in bins of $( Q^2 , s_1 , s_2 )$. 
The results obtained on the hadronic current enable us to
determine all sixteen structure functions $W_X$ differential
in the Dalitz plot. No significant non-axial-vector
contributions have been found. 

\section{ACKNOWLEDGEMENTS}
\label{s-ack}

We gratefully acknowledge the effort of the CESR staff in providing us with
excellent luminosity and running conditions.
J.R. Patterson and I.P.J. Shipsey thank the NYI program of the NSF, 
M. Selen thanks the PFF program of the NSF, 
M. Selen and H. Yamamoto thank the OJI program of DOE, 
J.R. Patterson, K. Honscheid, M. Selen and V. Sharma 
thank the A.P. Sloan Foundation, 
M. Selen and V. Sharma thank the Research Corporation, 
F. Blanc thanks the Swiss National Science Foundation, 
and H. Schwarthoff and E. von Toerne thank 
the Alexander von Humboldt Stiftung for support.  
This work was supported by the National Science Foundation, the
U.S. Department of Energy, and the Natural Sciences and Engineering Research 
Council of Canada.

\end{document}